\documentclass[journal]{IEEEtran}
\usepackage{amsmath,amssymb,amsfonts}
\usepackage{amsthm}
\usepackage{calligra}
\usepackage{algorithmicx}
\usepackage[linesnumbered,ruled,vlined]{algorithm2e}
\usepackage{bm}
\usepackage{cite}
\usepackage{hyperref}
\usepackage{enumerate}
\usepackage[table]{xcolor}
\usepackage{graphicx}
\usepackage{subfig}
\usepackage{epstopdf}
\usepackage{cite,url,epsfig}
\usepackage{epstopdf}
\usepackage{booktabs}
\usepackage{diagbox}
\usepackage{slashbox,multirow}

\begin{document}
	\title{Generative AI-enabled Blockchain Networks: Fundamentals, Applications, and Case Study}
	
	\author{ 
	Cong T. Nguyen, Yinqiu Liu, Hongyang Du, Dinh Thai Hoang, Dusit Niyato, Diep N. Nguyen, and Shiwen Mao 
		\thanks{Cong T. Nguyen is with the Institute of Fundamental and Applied Sciences, Duy Tan University, Vietnam. (e-mail: nguyenthanhcong21@duytan.edu.vn)
  
  Y.~Liu, H.~Du, and D. Niyato are with the School of Computer Science and Engineering, Nanyang
Technological University, Singapore (e-mail: yinqiu001@e.ntu.edu.sg, hongyang001@e.ntu.edu.sg, dniyato@ntu.edu.sg).
							
       Dinh Thai Hoang and Diep N. Nguyen are with the School of Electrical and Data Engineering, University of Technology Sydney, Australia (e-mail: Hoang.Dinh@uts.edu.au, Diep.Nguyen@uts.edu.au)
      
       S. Mao is with the Department of Electrical and Computer Engineering,
Auburn University, Auburn, USA (e-mail: smao@ieee.org)
       }
		}

	\maketitle
	
	\begin{abstract}
		Generative Artificial Intelligence (GAI) has recently emerged as a promising solution to address critical challenges of blockchain technology, including scalability, security, privacy, and interoperability. In this paper, we first introduce GAI techniques, outline their applications, and discuss existing solutions for integrating GAI into blockchains. Then, we discuss emerging solutions that demonstrate the effectiveness of GAI in addressing various challenges of blockchain, such as detecting unknown blockchain attacks and smart contract vulnerabilities, designing key secret sharing schemes, and enhancing privacy. Moreover, we present a case study to demonstrate that GAI, specifically the generative diffusion model, can be employed to optimize blockchain network performance metrics. Experimental results clearly show that, compared to a baseline traditional AI approach, the proposed generative diffusion model approach can converge faster, achieve higher rewards, and significantly improve the throughput and latency of the blockchain network. Additionally, we highlight future research directions for GAI in blockchain applications, including personalized GAI-enabled blockchains, GAI-blockchain synergy, and privacy and security considerations within blockchain ecosystems.		
		
	\end{abstract} 
	
	\begin{IEEEkeywords}
			Generative Artificial Intelligence, Blockchain, Variational Autoencoder, Generative Adversarial Network, Generative Diffusion Model, Large Language Model
	\end{IEEEkeywords}
	\IEEEpeerreviewmaketitle

	\section{Introduction}\label{Intro}
	Blockchain technology, renowned for its exceptional ability to maintain data integrity and immutability in decentralized settings, has been increasingly recognized as a crucial enabler for transparent data management. Fundamentally, a blockchain operates as a distributed ledger, where records are collectively maintained and shared across a peer-to-peer network. This technology leverages sophisticated cryptographic methods and consensus mechanisms to provide self-governance, security, transparency, and efficiency. Its applications are diverse and far-reaching, encompassing sectors such as finance and healthcare, and extending to innovative domains like the Metaverse and Web 3.0\cite{liu2020blockchain,xiao2020survey}. 
	
	While blockchain technology is innovative, it faces challenges such as scalability, security, privacy, and interoperability \cite{liu2020blockchain,xiao2020survey}. The integration of traditional Discriminative Artificial Intelligence (DAI) with blockchain shows great promise in addressing these issues~\cite{liu2020blockchain}. DAI can streamline blockchain operations in various ways. It can enhance scalability by compressing transaction data, refining consensus mechanism design, and optimizing network resource allocation. For security and accuracy, Natural Language Processing (NLP) techniques are instrumental in analyzing and verifying smart contracts, thereby preventing errors. Deep Learning (DL) significantly bolsters security as it can scrutinize transaction patterns and node information to spot and counteract malicious activities. Additionally, DL plays a vital role in preserving user privacy within blockchain networks because it can anonymize and consolidate transaction data and facilitate the processing of encrypted data without the need for decryption. In terms of interoperability, DAI aids in the development of secure and efficient cross-chain protocols and supports the processing of data at the semantic level, thereby enhancing the functionality and reach of blockchain technology.
	
While DAI techniques offer solutions to several blockchain challenges, they face notable limitations. Newer blockchain networks, often lacking extensive historical data, impede DAI's effectiveness in learning and preventing emerging threats. Additionally, DAI models trained on one type of blockchain, such as those using Proof-of-Work protocols, may not perform well on networks using different protocols like Proof-of-Stake or Proof-of-Authority. Moreover, DAI's inability to generate new content or adapt to novel scenarios limits its usefulness in applications like detecting new attacks or automating smart contract creation. Therefore, while DAI is beneficial in certain areas, its limitations highlight the need for more versatile technologies in blockchain development.
	
As AI technologies continue to evolve, Generative AI (GAI) has recently been emerging as a focal point, drawing even greater levels of attention. GAI techniques focus on generating new information, such as pictures, texts, sounds, videos, and system designs, by learning patterns and structures from existing data and autonomously producing new output~\cite{de2022deep}. Compared to DAI, GAI offers a distinct advantage thanks to its outstanding ability to generate data, coupled with its innate creativity and flexibility. Particularly, GAI can create realistic content by using latent vectors to represent given samples and learning their distribution. This allows GAI to overcome data scarcity by synthesizing new data \cite{du2023beyond}. 

	\begin{figure*}[t] 
		\centering
		\includegraphics[width=.95\textwidth ]{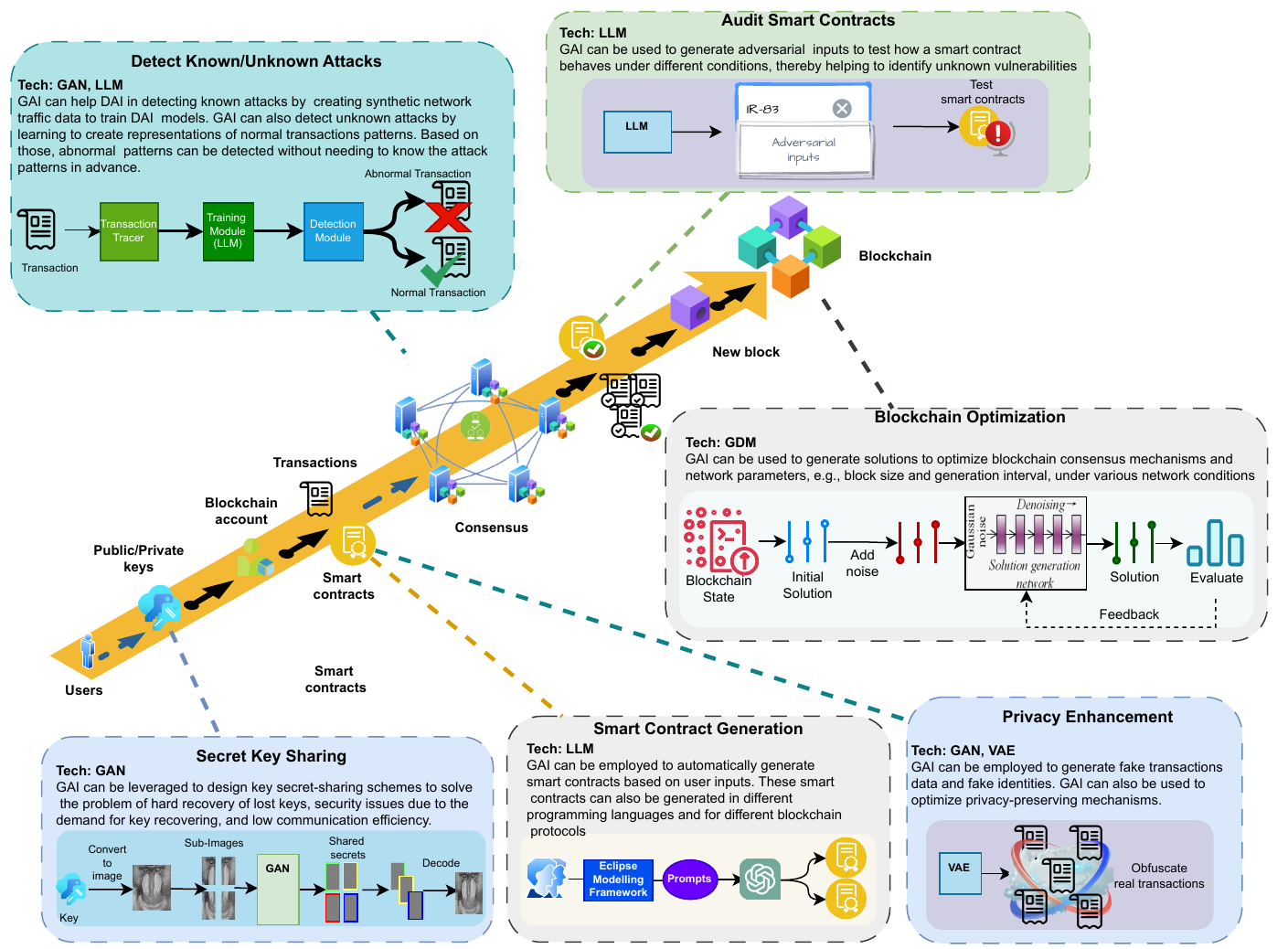}
		\caption{The schematic of GAI-enabled blockchain. 1) A user generates a public/private key pair to join a blockchain network. GAI can aid in key generation and sharing processes. 2) Once joined, the user can create transactions and smart contracts. GAI can automatically generate smart contracts. 3) Transactions and smart contracts are validated by the consensus mechanism. GAI can audit smart contracts and detect attacks from transactions. GAI can also be leveraged to optimize blockchain network parameters and consensus mechanisms. 4) Once validated, transactions and smart contracts are collected to create a new block to add to the chain. GAI also can generate fake transactions to obfuscate real transactions to improve privacy.}  \label{fig:system}
	\end{figure*}  
	
	Thanks to these advantages, GAI has the potential to address many challenges currently faced by DAI in blockchain networks, as illustrated in Fig.~\ref{fig:system}. Moreover, there are application scenarios in which GAI may be the sole viable solution. Particularly, GAI can be beneficial for the following scenarios:
	\begin{itemize}
		\item 	\textbf{Data augmentation for supporting DAI}: Data generated by GAI can be used to augment  DAI's training or to simulate and evaluate blockchain networks~\cite{ferrag2023generative}. 
		\item \textbf{Smart contract generation and vulnerabilities detection}: GAI can be used to generate adversarial inputs to test smart contracts, enabling it to effectively detect unknown vulnerabilities~\cite{david2023you}. Moreover, GAI can aid in automatic smart contract generation~\cite{petrovic2023model}.
		\item \textbf{Zero-day attack detection}: By generating transactions that mimic normal transactions for training, GAI can accurately detect unknown attacks via abnormal transaction patterns in blockchain networks~\cite{gai2023blockchain}.
		\item \textbf{Domain adaptation}: GAI can be utilized in cross-chain and cross-platform scenarios, such as creating new blockchain data based on existing blockchains with different applications/architecture or migrating existing networks into new protocols.
		\item  \textbf{Privacy enhancement}: GAI can be leveraged to generate fake transactions to obfuscate and anonymize user transaction history in blockchains.  
		\item 	\textbf{Scalability}: GAI can generate data to support the simulations and evaluations of newly designed consensus, cross-chain communication, and sharding mechanisms, thereby addressing scalability issues. 
        \item \textbf{Optimization}: GAI can learn to generate potential solutions for blockchain's optimization problems, e.g., determine block size and block time. 
	\end{itemize}

	Given the above potentials, this article provides a comprehensive exploration of how GAI can address the current challenges in blockchain, such as scalability, security, privacy, and interoperability. Particularly, we first introduce different types of GAI techniques and then summarize the potential applications and existing solutions for integrating GAI into blockchain networks. Moreover, we conduct a case study on blockchain design, focusing on how to leverage a GAI technique, namely Generative Diffusion Model (GDM), to optimize the blockchain network performance, e.g., throughput and latency. Simulation results show that compared to a baseline traditional AI approach, the proposed GDM approach can converge faster, achieve higher rewards, and significantly improve the throughput and latency of the blockchain network. Finally, we discuss potential future research directions of GAI applications for blockchain, including personalized GAI-enabled blockchain, GAI-blockchain synergy, and the privacy and security concerns of GAI applications in blockchain.

	\section{Overview Of AI-aided Blockchain Technology} 
	\label{sec:2}
	\subsection{Blockchain Fundamentals}
	
	Blockchain represents a novel paradigm for decentralized data management. A blockchain functions as a decentralized database, essentially a ledger, that shares records among participants within a peer-to-peer network. It relies on cryptographic hash functions, digital signatures, and distributed consensus mechanisms to ensure that once a record is added to the database, it cannot be changed without an agreement of other network participants. As a result, data stored in the blockchain can be verified without the need of a central authority~\cite{liu2020blockchain,xiao2020survey}.
	
	Transactions are the fundamental components of a blockchain network. They represent digital exchanges of assets or information between users, such as a transfer of network tokens, e.g., coins in cryptocurrency, among different users or an exchange of digital assets. Multiple transactions are bundled into a block, and the block is then added to an ever-growing sequence of blocks, i.e., a chain of blocks. The transactions and blocks are linked by hash pointers, such that any change in the transaction history can be immediately detected. To add new blocks to the chain, blockchain users participate in a consensus mechanism to preserve the network's security and integrity. Particularly, users in a blockchain network may exhibit various issues, including being faulty, engaging in malicious activities, or having inaccurate information. In such trustless environments, consensus mechanisms play a key role in ensuring that all users agree on the state of the network. For example, Proof of Work (PoW) and Proof of Stake (PoS) are two common consensus mechanisms. Participants, i.e., users, in a PoW-based blockchain network need to solve intensive computational puzzles to add new blocks to the chain. In contrast, participants in PoS are chosen to produce new blocks depending on the number of network tokens that they hold. These selection processes are necessary to ensure security and trust without the presence of central authorities~\cite{liu2020blockchain,xiao2020survey}. 
	
	\subsection{Challenges and Existing DAI Solutions}
	Despite its potential and numerous applications, blockchain technology also faces many challenges. To address the challenges, existing DAI solutions have been proposed.

	\subsubsection{Scalability}
	As the number of blockchain users is continuously growing, blockchain networks need to handle more and more transactions. This leads to serious scalability issues due to the trade-off between transaction throughput and network security in conventional blockchain networks. Particularly, increasing block size or reducing block time (i.e., average time to produce one block) can increase the transaction processing speed. However, it also leads to increasing risks of forks, attacks, or inconsistencies. DAI can be leveraged to address those challenges, such as using reinforcement learning to optimize consensus mechanisms~\cite{zou2023optimized} and resource allocation~\cite{liu2020blockchain}. However, acquiring an adequate amount of data to ensure the effectiveness of DAI training, especially labeled data, is not always practical.

	\subsubsection{Security}
	DAI has been widely adopted to address security issues in blockchain networks. For example, smart contracts can contain bugs, errors, or malicious codes that compromise their functionality and integrity. To address this issue, NLP can be employed to analyze and verify smart contract codes. Furthermore, DL can help to generate and encrypt digital signatures, thereby improving their resilience to attacks. Alternatively, DL techniques can be leveraged to detect and prevent fraud and attacks by analyzing blockchain transactions and node information~\cite{khoa2022deep}. Nevertheless, these DAI techniques often struggle when labeled data is limited. Moreover, they are not effective in detecting zero-day attacks and unknown vulnerabilities in smart contracts.
	
	\subsubsection{Privacy} 
	One of the main challenges of privacy is the trade-off between transparency and anonymity, as revealing too much or too little information can affect the trust and accountability of a blockchain network. To improve the privacy of blockchain networks, DAI solutions can be applied, such as using homomorphic encryption to perform direct computations on encrypted data and using federated learning to train models on distributed data without sharing them~\cite{jia2021blockchain}.

	\subsubsection{Interoperability} 
	As more blockchain applications emerge, the number of blockchain networks is rapidly increasing. However, due to the lack of common standards and protocols, these networks often cannot communicate with each other, which leads to serious interoperability issues. To improve the interoperability of blockchain networks, DAI techniques can be potential solutions, such as using ontology-based semantic web technologies to enable common understanding and representation of data across different blockchains. Moreover, transfer learning can be leveraged to enable cross-platform learning and adaptation. However, DAI techniques might not work well when two blockchain networks are employing different consensus mechanisms, e.g., PoS and PoW, or when they have different architectures, e.g., sharded and non-sharded blockchains.
	
From the above discussion, it is evident that DAI can be utilized to tackle various challenges in blockchain technology. However, it is worth noting that DAI has certain limitations. These include the reliance on labeled data, a lack of capability in detecting zero-day attacks and unknown vulnerabilities, and interoperability challenges while being applied across different consensus mechanisms and blockchain architectures.

	\section{Generative AI for Blockchain}
	\label{sec:3}
		In this section, we explore the potential of GAI to address the challenges of blockchain networks. Specifically, we first present the fundamentals of GAI and introduce four typical GAI models. Then, we discuss the existing GAI solutions to address various blockchain challenges, especially the ones that cannot be addressed by DAI. 
	\subsection{Fundamentals of Generative AI}
	With outstanding advantages in creativity and flexibility, GAI has become a promising solution to address the aforementioned challenges in blockchain technology. GAI focuses on creating new content and information based on training and user inputs. To this end, GAI, by using DL techniques and neural networks, can analyze patterns and structures in existing data. It then utilizes these learned characteristics to produce new data that closely resemble the original information. Moreover, by generating more content and learning from their own results, GAI can progressively enhance their performance and the quality of generated content~\cite{de2022deep}. 
	
	A key distinction between GAI and DAI lies in their approaches to data. Particularly, GAI aims to model the data's distribution, whereas DAI focuses on modeling the relationship between the data and its labels. For example, GAI can generate an image of a cat based on its learned knowledge of a typical cat's features. In contrast, DAI can classify an image as a cat or not based on its characteristics. Furthermore, GAI can utilize the learned distribution to produce new content, while DAI can only use the learned relationship for making predictions. As a result, GAI can learn to produce novel and diverse content that is related to but not limited by the training data. Comparisons of the integration of DAI and GAI with blockchain are summarized in Table~\ref{tab:tab1}.
	Next, we are going to discuss typical GAI models~\cite{xu2024unleashing} that have high potentials for blockchain networks.
	\subsubsection{Variational Autoencoder (VAE)} 
	A VAE comprises an encoder network, responsible for mapping input data into a latent space distribution, and a decoder network, which generates data samples from this distribution. This unique architecture enables VAE to learn compact and continuous data representations within a lower-dimensional latent space. As a result, VAE is a highly efficient solution for generating data based on long-term distributions, such as transaction history or smart contract usage.
	
	\subsubsection{Generative Adversarial Network (GAN)} A GAN consists of a generator neural network for creating synthetic data samples and a discriminator network that learns to distinguish real and fake data. Through adversarial training of both networks, GAN excels at generating high-quality data, which is essential for training blockchain attack detection systems or for simulation and evaluation purposes~\cite{ferrag2023generative}.
	
	\subsubsection{Generative Diffusion Model (GDM)} A GDM iteratively adds noise to an initial data point and then denoises it, gradually converging to the desired data distribution. Thanks to their ability to produce realistic and diverse data samples, GDMs can be applied to generate high-quality data for evaluation or generate solutions to blockchain optimization problems.
	
	\subsubsection{Large Language Model  (LLM)} An LLM is an AI model designed for natural language understanding and generation tasks. These models are built on deep neural networks with millions to billions of parameters, enabling them to capture and generate human-like text across a wide range of languages and topics. As a result, LLMs can be especially useful for understanding and generating text-based data in the blockchain, such as smart contract code~\cite{david2023you,petrovic2023model}.
	
	\begin{table*}[]
		\caption{Comparisons of the applications of DAI and GAI in blockchain networks}
		\label{tab:tab1}
		\begin{tabular}{ll|l|l|}
			\cline{3-4}
			&                                                                               & \multicolumn{1}{c|}{\textbf{Discriminative Artificial Intelligence (DAI)}}                                                                                                                                                                                             & \multicolumn{1}{c|}{\textbf{Generative Artificial Intelligence (GAI)}}                                                                                                                                                                                                                                                                   \\ \hline
			\multicolumn{1}{|l|}{\multirow{10}{*}{\rotatebox[origin=c]{90}{\textbf{Characteristics}}}} & \begin{tabular}[c]{@{}l@{}}Implementation \\ approach\end{tabular}            & \begin{tabular}[c]{@{}l@{}}-Utilize available blockchain data such as transaction\\ history and smart contracts to train models\end{tabular}                                                                                   & \begin{tabular}[c]{@{}l@{}}-Generate high-quality synthetic blockchain data and contents\end{tabular}                                                                                                                                                                                             \\ \cline{2-4} 
			\multicolumn{1}{|l|}{}                                          & \begin{tabular}[c]{@{}l@{}}Data \\ availability\end{tabular}                  & \begin{tabular}[c]{@{}l@{}}-Require sufficient high-quality data \\ -Might not work well in new blockchain networks \\ with little historical data\end{tabular}                                                             & \begin{tabular}[c]{@{}l@{}}-Do not require a lot of data \\ -Create data, e.g., using GAN and VAE, to support DAI training\end{tabular}                                                                                                                                                                                         \\ \cline{2-4} 
			\multicolumn{1}{|l|}{}                                          & \begin{tabular}[c]{@{}l@{}}Domain \\ adaptation\end{tabular}                  & \begin{tabular}[c]{@{}l@{}}-Cannot adapt to different smart contract \\ languages or blockchain protocols\end{tabular}                                                                                                  & \begin{tabular}[c]{@{}l@{}}-Adapt to different smart contract languages and blockchain \\protocols\end{tabular}                                                                                                                                                                            \\ \cline{2-4} 
			\multicolumn{1}{|l|}{}                                          & Uncreativity                                                                  & \begin{tabular}[c]{@{}l@{}}-Struggle to produce new content \\ -Cannot detect unknown blockchain attacks and smart \\ contract vulnerabilities\end{tabular}                                                                      & \begin{tabular}[c]{@{}l@{}}-Create new contents, e.g., automatically create smart contracts\\ using LLMs \\ -Work well with unknown scenarios, e.g., detect  new attacks and\\ vulnerabilities\end{tabular}                                                                                                                                           \\ \hline
			\multicolumn{1}{|l|}{\multirow{15}{*}{\rotatebox[origin=c]{90}{\textbf{Applications}}}}    & \begin{tabular}[c]{@{}l@{}}Blockchain \\ network \\ optimization\end{tabular} & \begin{tabular}[c]{@{}l@{}}-Use reinforcement learning for resource allocation\\ -Optimize consensus mechanisms\\ \textbf{Disadvantages}: Might not work well in different \\ network conditions and blockchain protocols\\ due to a lack of adaptability\end{tabular} & \begin{tabular}[c]{@{}l@{}}-Use GAN, VAE, and LLM to generate transactions, traffic data,\\ and smart contracts to support simulation and optimization\\ -Use GDM to optimize blockchain network design\\ \textbf{Advantages}: Data augmentation. Can adapt to different network  \\conditions and blockchain protocols\end{tabular}    \\ \cline{2-4} 
			\multicolumn{1}{|l|}{}                                          & Attack detection                                                              & \begin{tabular}[c]{@{}l@{}}-Analyze traffic and transactions to detect attacks\\ \textbf{Disadvantages}: Can only detect known attacks\end{tabular}                                                                                    & \begin{tabular}[c]{@{}l@{}}-Mimic normal traffic and transactions patterns. Flag abnormal \\patterns\\ \textbf{Advantages}: LLM can detect unknown attacks, e.g., recent \\attacks on Beanstalk and Revest~\cite{gai2023blockchain}. \end{tabular}                                                                                                                                         \\ \cline{2-4} 
			\multicolumn{1}{|l|}{}                                          & Smart contract design                                                              & \begin{tabular}[c]{@{}l@{}}-Analyze smart contract code to detect vulnerabilities\\ \textbf{Disadvantages}: Can only detect known vulnerabilities\\ Cannot generate smart contract\end{tabular}                                       & \begin{tabular}[c]{@{}l@{}}-Understand smart contract codes\\ -Use LLM to automatically generate smart contracts\\ \textbf{Advantages}: Can detect unknown vulnerabilities \\ Can generate smart contracts in different languages, e.g., Solidity\\ and DAML\end{tabular}                                                                 \\ \cline{2-4} 
			\multicolumn{1}{|l|}{}                                          & Privacy                                                                       & \begin{tabular}[c]{@{}l@{}}-Assist in encryption and anonymization\\ \textbf{Disadvantages}: Rely on quality and quantity of \\ training data\end{tabular}                                                                         & \begin{tabular}[c]{@{}l@{}}-Use GAN and VAE to generate fake transactions to obfuscate\\ real transactions\\ \textbf{Advantages}: Do not require a lot of data. Can generate \\ high-quality fake data\end{tabular}                                                                                                              \\ \cline{2-4} 
			\multicolumn{1}{|l|}{}                                          & Interoperability                                                              & \begin{tabular}[c]{@{}l@{}}-Use transfer learning to enable cross-platform \\ learning and adaptation\\ \textbf{Disadvantages}: Cannot adapt to significant\\ changes/differences in networks and protocols\end{tabular}                                  & \begin{tabular}[c]{@{}l@{}}-Generate synthetic representations of multiple blockchain networks \\ -Use GAN to generate new blockchain data in different protocols\\ \textbf{Advantages}: Can adapt to different protocols and settings \\ Can help to migrate blockchain data to different networks and \\protocols\end{tabular} \\ \hline
		\end{tabular}
	\end{table*}
	\subsection{Generative AI for Blockchain}
	\subsubsection{Challenges that GAI can improve over DAI}
	\paragraph{Detecting known attacks} GAI can be used to help DAI to detect anomalies in blockchain networks. For example, in~\cite{ferrag2023generative}, a GAN model is used to create synthetic data from a real network traffic log dataset. Then, the newly created data is used to train a Transformer-based model to detect and prevent cyberattacks. After training, the Transformer-based model can analyze network traffic and detect potential cyberattacks. Simulation results show that the proposed approach can detect cyberattacks with over 95\% accuracy. 
	\paragraph{Audit smart contracts} Both DAI and GAI can be applied to audit smart contracts in different ways. DAI can be trained to identify specific vulnerabilities, e.g., frontrunning, backrunning, and sandwiching~\cite{david2023you}, in smart contracts. Similar to the intrusion detection cases, DAI is only effective for detecting known vulnerabilities or patterns of risky behavior in contract code. On the other hand, GAI can be used to generate adversarial inputs to test how a smart contract behaves under different conditions. This can help to detect unknown vulnerabilities in smart contract code, exceeding the capability of DAI. For example, the authors in~\cite{david2023you} conduct experiments on a benchmark dataset of 10,000 existing smart contracts and compare the performance of two LLMs, e.g., GPT4 and Claude, and a random baseline model. The results show that LLMs outperformed the random model by 20\% in terms of attack detection accuracy, demonstrating the potential of LLMs for enhancing security analysis and improving the efficiency of the auditing process.

	\subsubsection{Unique challenges that only GAI can address}
	\paragraph{Detect unknown attacks} In addtion to enhancing the performance of DAI in intrusion detection, GAI has a unique ability to accurately detect zero-day attacks. Particularly, although DAI is often employed for anomaly detection, it only performs well if the characteristics of attacks are known and there is a clear distinction between normal and malicious activities in the dataset. In contrast, GAI can deal with the cases where attack patterns are not well-defined or are evolving over time, as GAI can capture a broader range of anomalies. For example, a GAI-based blockchain intrusion detection system, namely BLOCKGPT, is proposed in~\cite{gai2023blockchain}. BLOCKGPT can detect two real-world attacks, i.e., Beanstalk and Revest, which caused more than \$80 million loss, based on analyzing the transaction traces. Moreover, experimental results show that BLOCKGPT can process, on average, 2284 transactions per second, and it can detect twice as many new attacks as those of the baseline methods. 
	
	\paragraph{Generate smart contracts} GAI techniques, such as GANs and LLMs, can be applied for smart contract generation by learning and simulating the patterns and logic found in existing smart contracts. For example, GAI can generate code that adheres to the syntax and semantics of smart contract programming languages such as Solidity (used in Ethereum). It can assist developers by automatically generating code templates, suggesting contract structures, or even proposing entire contracts based on high-level descriptions or requirements. For example, the authors in~\cite{petrovic2023model} investigate the potential of ChatGPT for smart contract generation. The authors develop an Eclipse Modeling Framework~\cite{petrovic2023model} to translate users' input parameters (e.g., smart contract participants and transactions receiver) into prompts. These prompts are then fed into ChatGPT to automatically generate smart contracts in two languages, e.g., Solidity and DAML. Moreover, smart contracts can be generated for different blockchain networks employing different protocols by slightly adjusting the prompts. Additionally, GAI can detect potential vulnerabilities by simulating contract interactions and avoiding those during the generation process.
	\paragraph{Optimize blockchain network designs} GAI can be applied to optimize blockchain designs by creating synthetic workloads and transaction patterns that mimic real-world usage scenarios. Based on those, blockchain developers and network administrators can simulate different resource allocation strategies and optimize them for high efficiency and performance. Alternatively, GAI techniques such as GDM can be leveraged to directly generate potential solutions to resource allocation problems. In Section~\ref{case} of this paper, we will present a case study to demonstrate the effectiveness of this approach.  

	\paragraph{Design key secret-sharing schemes} GAI can support to design key secret-sharing schemes to solve the problem of hard recovery of lost keys, security issues due to the demand for recovering the private key, and low key communication efficiency in blockchain. As shown in \cite{zheng2020gan}, the secret-sharing process can be treated as a classification problem of images. Particularly, a private key can be converted into an image. Then, the image is segmented into sub-images. These are then used to train the generator network of the GAN to generate shared secret subimages from noise, while the discriminator network determines whether the generated subimages are consistent or similar to the original subimages. When the process is finished, the shared secret subimages created by the generator can be decoded to construct the original image. Simulation results show that the proposed scheme can recover the original image with the highest quality, e.g., 19\% higher in terms of peak-signal-to-noise-ratio (PSNR), compared to other baseline methods.

	\paragraph{Enhance privacy} GAI offers unique advantages in improving blockchain privacy. Generative models like VAEs and GANs can generate synthetic data that closely mimic real blockchain transactions and activities. This synthetic data can be used to obscure or mask sensitive information, making it challenging for adversaries to trace or de-anonymize users on the blockchain. Moreover, GAI can support the encryption of transactions or identity data, e.g., applying the scheme proposed in \cite{zheng2020gan} for transactions and identity data. The major use cases of GAI in blockchain are summarized in Table~\ref{tab:tab2}.
	\begin{table*}[!]
		\centering
		\caption{Summary of GAI approaches for blockchain.} 
		\begin{tabular}{|>{\raggedright\arraybackslash}m{2.5cm}|>{\raggedright\arraybackslash}m{1.5cm}|>{\raggedright\arraybackslash}m{5.5cm}|>{\raggedright\arraybackslash}m{5.5cm}|}
			\hline 
			{\centering\arraybackslash}{\textbf{}} &
			{\centering\arraybackslash}{\textbf{Technique}} &
			{\centering\arraybackslash}{\textbf{How it works}}&
			{\centering\arraybackslash}{\textbf{Effectiveness}}
			\\
			\hline 
			\hline
			\textbf{Detect known attacks}          &GAN   & GAN generate synthetic data from real network traffic log to help training attack detection model   & Detect cyberattacks with more than 95\% accuracy~\cite{ferrag2023generative}         \\ 
			\hline
			\textbf{Smart contract audit}     &LLM       & Use LLM to analyze smart contract codes to detect vulnerabilities            & Detect twice as many unknown vulnerabilities as baseline methods.~\cite{david2023you}  \\ 
			\hline
			
			\textbf{Detecting unknown Attacks}     &LLM    &     Learn to mimic real tracing representations of transactions. Flag abnormal transaction traces   & Detect twice as many new attacks as baseline methods~\cite{gai2023blockchain}                 \\ 
			\hline
			\textbf{Smart contract generation}     &LLM    &     Design prompts to ask LLMs to generate smart contracts        & Generate smart contracts in multiple languages and for different protocols~\cite{petrovic2023model}     \\ 
			\hline
			\textbf{Blockchain optimization}     &GDM    &     Train GDM to generate solutions for optimizing blockchain designs       & Converge faster and perform better than DRL (as shown in Section~\ref{case})        \\ 
			\hline
			\textbf{Key secret-sharing scheme}     &GAN    &     Convert secret key to image and learn to generate secret shares based on images       & Secret recovery with 19\% higher PSNR~\cite{zheng2020gan}          \\ 
			\hline
			\textbf{Privacy enhancement}     &GAN, VAE    &  Generate fake transactions data and fake identities. Help in transaction and identity encryption, e.g., apply the scheme in \cite{zheng2020gan}        &   Obfuscate real transactions and identities             \\ 
			\hline
		\end{tabular}
		\label{tab:tab2}
	\end{table*}

	
	
	
	
	
	
	\section{Case Study: Diffusion Model-based Blockchain Design}
	\label{case}
	In this case study, we leverage GDM to optimize a blockchain system, showing how GAI can assist in optimizing a blockchain's performance.
	\begin{figure*}[tbp]
		\centerline{\includegraphics[width=1.75\columnwidth]{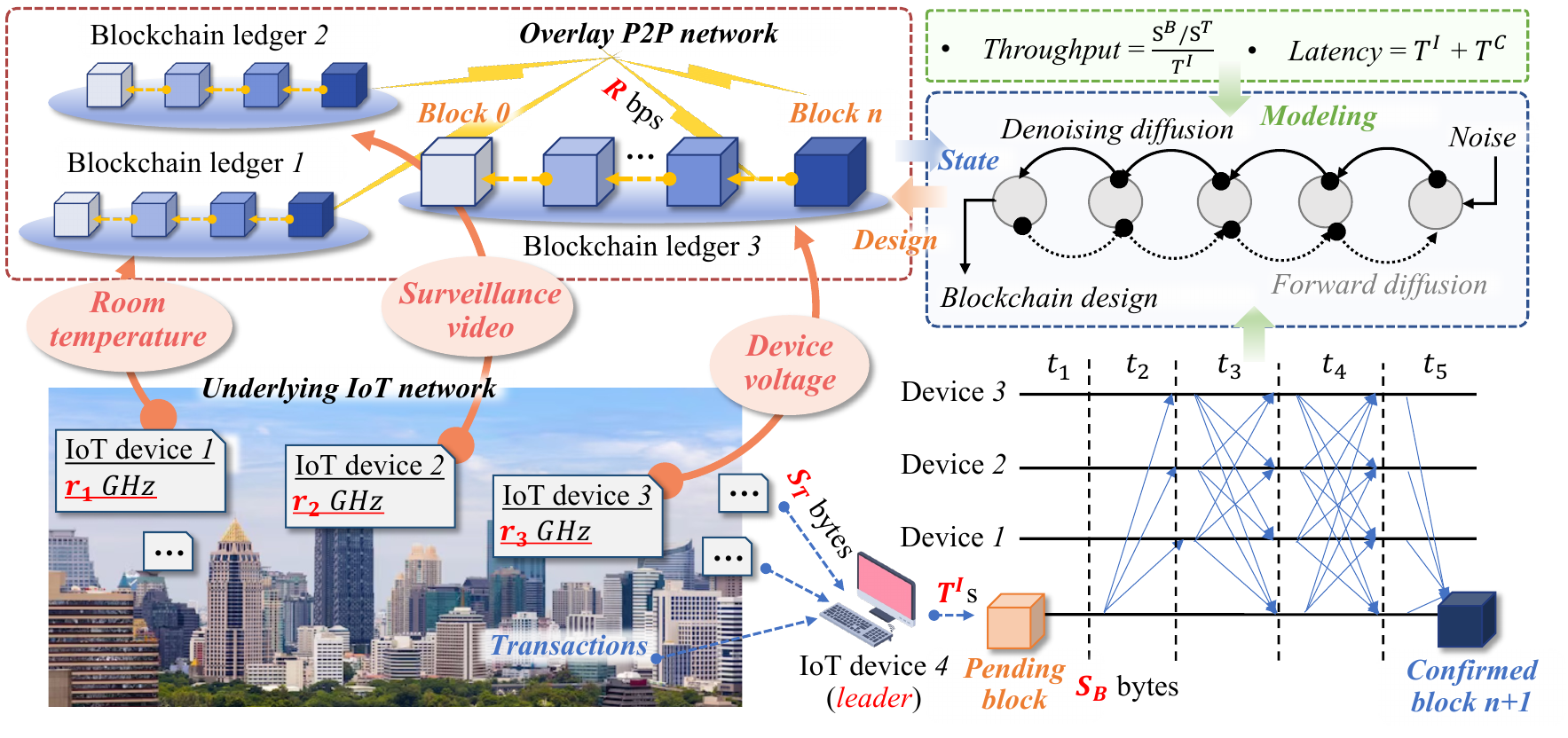}}
		\caption{The model of IoT-orient blockchain system. Note that in PBFT, each node performs two operations for message validation, namely signature validation and message endorsement. The corresponding computation complexity is denoted by $C_S$ and $C_E$, respectively. According to \cite{Blockchain2}, each node will perform $1$ and $2 + 4(K + f - 1)$ times of signature validation and message endorsement, respectively, where $f$ means the number of malicious block producers. Accordingly, $T^V$ = $\frac{C_S + [2+4(K+f-1)]C_E}{R}$. Since PBFT contains five rounds of broadcast, $T^B$ = $5 \frac{S_B}{R}$. The optimization goal is to maximize $\alpha \cdot \textit{throughput} + \beta \cdot \textit{latency}$, with the constraint that the latency should be less than the user threshold.} 
		\label{extraction}
	\end{figure*}
	
	\subsection{System Model}
	Our case study models a consortium blockchain for Internet of Things (IoT) data transmission, with a focus on optimizing blockchain performance in this context. We consider $N$ IoT devices with heterogeneous computational capabilities, forming a network where each data transmission between devices generates a blockchain transaction. To ensure an immutable and traceable transaction history, we select $K$ nodes (where $K \leq N$) as block producers, responsible for executing a consensus mechanism to form and validate blocks, thereby mitigating the risk of single-point failures \cite{Blockchain}.
	This model is implemented using open-source platforms like Hyperledger Fabric and Ethereum \cite{Blockchain}, covering the necessary data structures and operational workflows for block producers. We then adopt Practical Byzantine Fault Tolerance (PBFT) \cite{Blockchain}, a widely used consensus mechanism in consortium blockchains. PBFT involves a leader block producer generating a pending block at regular intervals, followed by a multi-stage voting process for block validation.
	

	\subsection{Problem Formulation}
	The advancement of lightweight and mining-free blockchains facilitates blockchain deployment on resource-constrained IoT devices. That said, the selection of block producers is important since IoT nodes exhibit heterogeneity in terms of physical resources. Only by selecting appropriate block producers with sufficient resources and configuring the suitable PBFT settings accordingly can maximize the blockchain performance in the given scenario. Therefore, this case study aims to provide a GAI-empowered approach for customized blockchain design. To enhance the performance of the considered blockchain, we focus on allocating key resources, namely computational power, storage capacity, and bandwidth. Accordingly, we formulate an optimization problem to fine-tune block producer selection, block size, and block time, as illustrated in Fig.~\ref{extraction}.
	
	Particularly, we first identify two Key Performance Indicators (KPIs) for blockchain performance: throughput and confirmation latency. Throughput represents the rate at which transactions are recorded on the blockchain, and confirmation latency refers to the time taken for a transaction to be fully confirmed by all block producers. As depicted in Fig.~\ref{extraction}, these KPIs are integrated using a linear function with adjustable weights, allowing us to assess the combined effect of throughput and latency on overall performance.
	
	As shown in Fig.~\ref{extraction}, throughput is affected by the block size $S^B$, the block time $T^I$, and the average size of one transaction $S^T$. Confirmation latency comprises the block time $T^I$ and block confirmation time $T^C$. Moreover, the duration of $T^C$ depends on the validation latency $T_V$ and broadcast time $T_B$, both of which depend on the selection of block producers. As a result, to optimize the KPIs, we need to find the optimal selection of block producers, as well as the optimal values for $S^B$ and $T^I$. Additionally, the primary constraint is controlling confirmation latency under a predetermined limit, as excessive delays can lead users to perceive transaction processing as unsuccessful.

	\subsection{Proposed GDM Approach}
	To address the above optimization problem and design high-performance blockchains, we present a GDM-based solution. 
	\begin{itemize}
		\item \textbf{Conditions for guiding the denoising process}: The condition space describes the blockchain network and is defined as $[\{r_1, \dots, r_N\}, R, S_T, C_S, C_E]$, where $\{r_1, \dots, r_N\}$ represent the computation resources of $N$ IoT devices; $R$ means the network bandwidth; $S_T$ represents the average size of one transaction; $C_S$ and $C_E$ means the computation complexity for verifying one signature and generating one endorsement, respectively. 
		\item \textbf{Generated solution}: The generated solution takes the form $[\{s_1, \dots, s_N\}, S_B, T_I]$. Particularly, $\{s_1, \dots, s_N\}$ indicate the scores of $N$ IoT devices, respectively, and the $K$ candidates with the highest scores are selected as the block producers. $S_B$ and $T_I$ represent the block size and block time, respectively\footnote{The units of these factors are shown in Fig.~\ref{extraction}.}.
		\item \textbf{Reward for training}: The reward takes the value of blockchain performance if the latency constraint is satisfied. Otherwise, the reward is set to \textbf{-500} for punishment.
	\end{itemize}
	Specifically, the GDM is trained on blockchain network data, learning to correlate specific conditions with generated solutions that maximize rewards. 
 This involves iterative simulations where the GDM, given a condition, generates a potential solution and evaluates the outcome based on the defined reward. Through this process, the denoising process is trained to generate the solution that can maximize the reward~\cite{du2023beyond}, identifying the most effective combination of block producers, block sizes, and block time that yields the highest performance under varying network conditions. 
	
	\begin{figure}[tbp]
		\centerline{\includegraphics[width=0.9\columnwidth]{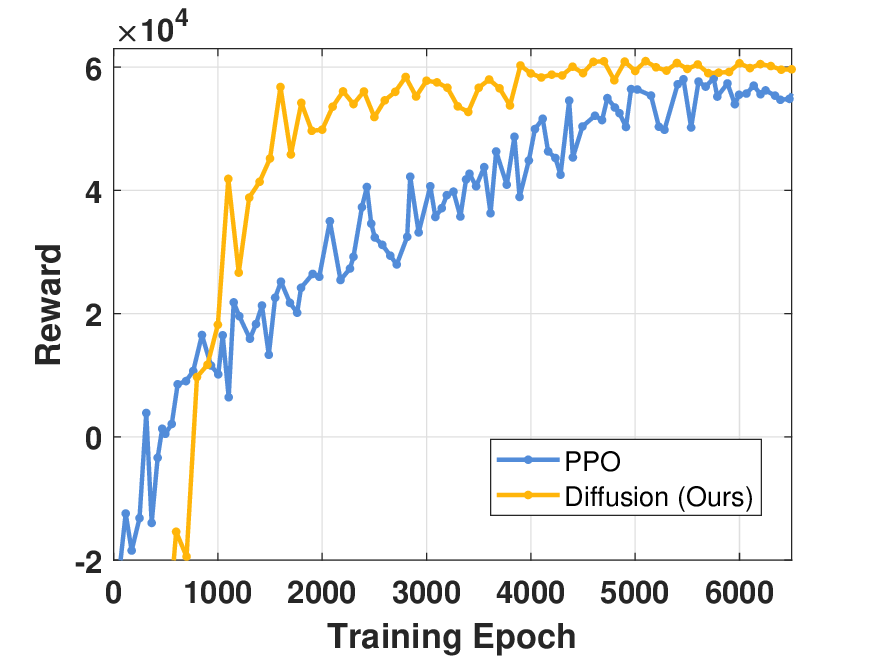}}
		\caption{The training curves of GDM and PPO.}
		\label{extraction2}
	\end{figure}
	\begin{figure}[tbp]
		\centerline{\includegraphics[width=0.8\columnwidth]{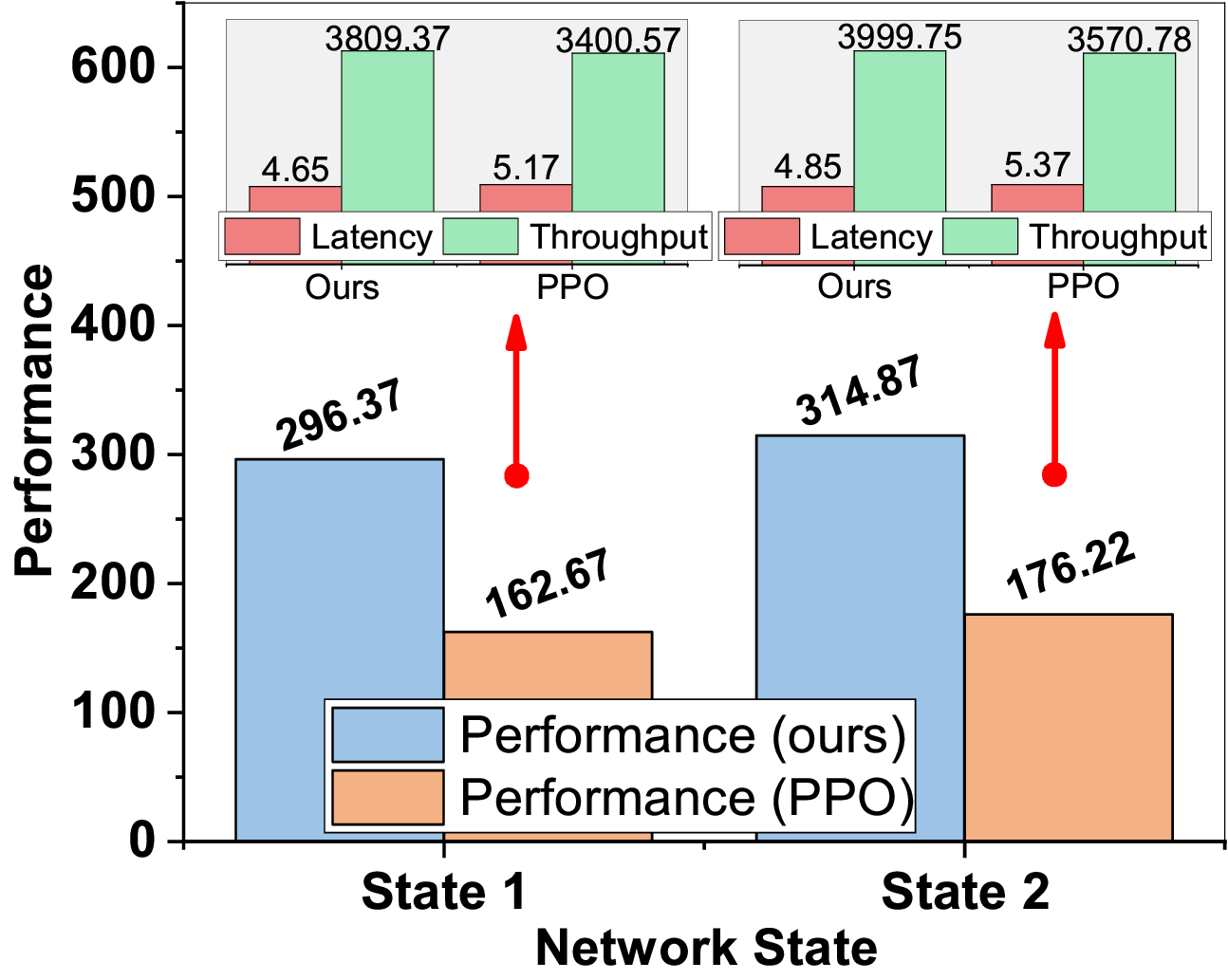}}
		\caption{The performance of blockchains designed by diffusion and PPO. Note that the two subfigures show the throughput and latency, whose units are \textit{transactions per second} and \textit{second}, respectively. The states 1 and 2 are [{12, 30, 15, 34, 15, 8}, 5500000, 210, 0.025, 0.01] and [{10, 28, 17, 32, 13, 10}, 5000000, 200, 0.02, 0.015], respectively.}
		\label{extraction3}
	\end{figure}

	\subsection{Simulation Results}
	We conduct simulations to evaluate and compare the effectiveness of the proposed GDM approach with a conventional Deep Reinforcement Learning (DRL) algorithm, particularly the Proximal Policy Optimization (PPO). Specifically, we consider a system with six IoT devices. According to PBFT theory \cite{Blockchain}, 3$f$ + 1 participants are required for defending $f$ attackers. For simplicity, we assume one attacker exists in the system. Accordingly, four block producers are required. As illustrated in Fig.~\ref{extraction2}, GDM exhibits a notably faster convergence rate than PPO. Specifically, our proposed approach converges approximately 1.5 times faster than PPO (around 4,000 epoch vs 6,000 epoch of the PPO). This accelerated convergence underscores the GDM's effectiveness in rapidly adapting to the complex demands of blockchain networks, which is particularly beneficial for applications requiring quick adjustments and low block times. Furthermore, our approach yields an approximately 8\% higher reward than PPO, demonstrating its superior optimization capability.
	
	Apart from the observed faster convergence and higher reward, GDM also outperforms PPO in terms of designing blockchains according to the specific condition state. As depicted in Fig.~\ref{extraction3}, GDM can increase the throughput by over 400 TPS. Meanwhile, the confirmation latency is slightly decreased. This performance edge can be attributed to the model's refined capability to balance block size, block time, block producer selection, and resource allocation, thus optimizing network resource utilization.
	
	\section{Future Directions}
	\label{sec:5}
		\subsection{Personalized Generative AI-enabled Blockchain}
	Personalized generative AI in the context of blockchain is a promising research direction that focuses on tailoring data generation to individual preferences. This approach can offer more effective and personalized solutions for individual users. For example, to generate artificial transactions to improve privacy, GAI needs to be trained with personalized transaction data to closely mimic the real transactions that the user often makes. To this end, techniques such as federated learning or meta-learning can be combined with GAI. However, this might also open new attack surfaces, e.g., sensitive transaction information leaks, that need to be investigated.
	
	\subsection{Privacy and Security}
	Privacy and security have always been the primary concerns in blockchain networks. As discussed above, although the integration of GAI in blockchain networks can address many security and privacy challenges, it might also create additional vulnerabilities. For example, personalized GAI might require access to sensitive personal data, and thus, managing data access is important. Moreover, if GAI services are deployed on smart contracts, there might be vulnerabilities that an adversary can exploit. Furthermore, GAI might be manipulated to create harmful content or malicious data to temper with intrusion detection or privacy-preserving mechanisms' training.    
 
	\subsection{GAI-Blockchain Synergy}
As discussed above, GAI offers promising solutions to address various challenges that blockchain is facing. On the other hand, blockchain can significantly enhance the privacy, security, and trustworthiness of GAI models and their training processes. Therefore, the collaboration between GAI and blockchain can establish a continuous cycle of improvement, allowing both technologies to mutually benefit and advance. For example, consider a decentralized platform for crowdsourcing where blockchain serves as a database to record user contributions. In this system, blockchain plays a key role in ensuring the immutability and transparency of the contribution data that GAI is trained on. Meanwhile, GAI can be employed to monitor the transaction history to detect abnormal patterns, e.g., fraudulent records of contributions. Moreover, GAI can also be trained collectively by users (as a crowdsourcing task) of the platform. As a result, GAI and blockchain complement each other, creating a resilient system where the strengths of each technology are utilized to enhance the overall integrity and security.
	
	\section{Conclusion}
	\label{sec:6}
	In this paper, we have explored GAI's potential to address various challenges of blockchain technology. Particularly, we have introduced fundamental concepts of blockchain technology and GAI techniques. Moreover, we have outlined and discussed the existing and potential applications of GAI in blockchain. Following this, we have demonstrated via a case study how the GDM technique can be leveraged to optimize blockchain consensus mechanisms and network parameters. Experiment results have shown that the GDM technique can converge faster, achieve higher rewards, and significantly improve the throughput and latency of the blockchain network compared to the traditional DAI approach. Finally, we have discussed potential research directions in the applications of GAI for blockchain technology.	
	
	\bibliographystyle{IEEEtran}
	\bibliography{ref}

\begin{thebibliography}{10}
\providecommand{\url}[1]{#1}
\csname url@samestyle\endcsname
\providecommand{\newblock}{\relax}
\providecommand{\bibinfo}[2]{#2}
\providecommand{\BIBentrySTDinterwordspacing}{\spaceskip=0pt\relax}
\providecommand{\BIBentryALTinterwordstretchfactor}{4}
\providecommand{\BIBentryALTinterwordspacing}{\spaceskip=\fontdimen2\font plus
\BIBentryALTinterwordstretchfactor\fontdimen3\font minus
  \fontdimen4\font\relax}
\providecommand{\BIBforeignlanguage}[2]{{%
\expandafter\ifx\csname l@#1\endcsname\relax
\typeout{** WARNING: IEEEtran.bst: No hyphenation pattern has been}%
\typeout{** loaded for the language `#1'. Using the pattern for}%
\typeout{** the default language instead.}%
\else
\language=\csname l@#1\endcsname
\fi
#2}}
\providecommand{\BIBdecl}{\relax}
\BIBdecl

\bibitem{liu2020blockchain}
Y.~Liu, F.~R. Yu, X.~Li, H.~Ji, and V.~C. Leung, ``Blockchain and machine
  learning for communications and networking systems,'' \emph{IEEE
  Communications Surveys \& Tutorials}, vol.~22, no.~2, pp. 1392--1431, 2020.

\bibitem{xiao2020survey}
Y.~Xiao, N.~Zhang, W.~Lou, and Y.~T. Hou, ``A survey of distributed consensus
  protocols for blockchain networks,'' \emph{IEEE Communications Surveys \&
  Tutorials}, vol.~22, no.~2, pp. 1432--1465, 2020.

\bibitem{de2022deep}
S.~De, M.~Bermudez-Edo, H.~Xu, and Z.~Cai, ``Deep generative models in the
  industrial internet of things: a survey,'' \emph{IEEE Transactions on
  Industrial Informatics}, vol.~18, no.~9, pp. 5728--5737, 2022.

\bibitem{du2023beyond}
H.~Du, R.~Zhang, Y.~Liu, J.~Wang, Y.~Lin, Z.~Li, D.~Niyato, J.~Kang, Z.~Xiong,
  S.~Cui \emph{et~al.}, ``Beyond deep reinforcement learning: A tutorial on
  generative diffusion models in network optimization,'' \emph{arXiv preprint
  arXiv:2308.05384}, 2023.

\bibitem{ferrag2023generative}
M.~A. Ferrag, M.~Debbah, and M.~Al-Hawawreh, ``Generative ai for cyber
  threat-hunting in 6g-enabled iot networks,'' \emph{arXiv preprint
  arXiv:2303.11751}, 2023.

\bibitem{david2023you}
I.~David, L.~Zhou, K.~Qin, D.~Song, L.~Cavallaro, and A.~Gervais, ``Do you
  still need a manual smart contract audit?'' \emph{arXiv preprint
  arXiv:2306.12338}, 2023.

\bibitem{petrovic2023model}
N.~Petrovi{\'c} and I.~Al-Azzoni, ``Model-driven smart contract generation
  leveraging chatgpt,'' in \emph{International Conference On Systems
  Engineering}.\hskip 1em plus 0.5em minus 0.4em\relax Springer, 2023, pp.
  387--396.

\bibitem{gai2023blockchain}
Y.~Gai, L.~Zhou, K.~Qin, D.~Song, and A.~Gervais, ``Blockchain large language
  models,'' \emph{arXiv preprint arXiv:2304.12749}, 2023.

\bibitem{zou2023optimized}
Y.~Zou, Z.~Jin, Y.~Zheng, D.~Yu, and T.~Lan, ``Optimized consensus for
  blockchain in internet of things networks via reinforcement learning,''
  \emph{Tsinghua Science and Technology}, vol.~28, no.~6, pp. 1009--1022, 2023.

\bibitem{khoa2022deep}
T.~V. Khoa, D.~T. Hoang, N.~L. Trung, C.~T. Nguyen, T.~T.~T. Quynh, D.~N.
  Nguyen, N.~V. Ha, and E.~Dutkiewicz, ``Deep transfer learning: A novel
  collaborative learning model for cyberattack detection systems in iot
  networks,'' \emph{IEEE Internet of Things Journal}, 2022.

\bibitem{jia2021blockchain}
B.~Jia, X.~Zhang, J.~Liu, Y.~Zhang, K.~Huang, and Y.~Liang,
  ``Blockchain-enabled federated learning data protection aggregation scheme
  with differential privacy and homomorphic encryption in iiot,'' \emph{IEEE
  Transactions on Industrial Informatics}, vol.~18, no.~6, pp. 4049--4058,
  2021.

\bibitem{xu2024unleashing}
M.~Xu \emph{et~al.}, ``Unleashing the power of edge-cloud generative ai in
  mobile networks: A survey of aigc services,'' \emph{IEEE Communications
  Surveys \& Tutorials}, 2024.

\bibitem{zheng2020gan}
W.~Zheng, K.~Wang, and F.-Y. Wang, ``Gan-based key secret-sharing scheme in
  blockchain,'' \emph{IEEE Transactions on Cybernetics}, vol.~51, no.~1, pp.
  393--404, 2020.

\bibitem{Blockchain2}
M.~Keshk, B.~Turnbull, N.~Moustafa, D.~Vatsalan, and K.-K.~R. Choo, ``A
  privacy-preserving-framework-based blockchain and deep learning for
  protecting smart power networks,'' \emph{IEEE Transactions on Industrial
  Informatics}, vol.~16, no.~8, pp. 5110--5118, 2019.

\bibitem{Blockchain}
Y.~Liu, K.~Qian, K.~Wang, and L.~He, ``Effective scaling of blockchain beyond
  consensus innovations and moore’s law: Challenges and opportunities,''
  \emph{IEEE Systems Journal}, vol.~16, no.~1, pp. 1424--1435, Mar. 2022.

\end{thebibliography}
\end{document}